# Modeling of multiple three phase contact lines of liquid droplets on geometrically patterned surfaces: continuum and mesoscopic analysis

Nikolaos T. Chamakos[a], Michail E. Kavousanakis[b] and Athanasios G. Papathanasiou[*c]

School of Chemical Engineering, National Technical University of Athens, 15780, Greece.
[a] E-mail: nhamakos@mail.ntua.gr. [b] E-mail: mihkavus@chemeng.ntua.gr.
[*c] Corresponding author. E-mail: pathan@chemeng.ntua.gr; Web: http://www.chemeng.ntua.gr/people/pathan.
Fax: +30 210772 3298; Tel: +30 210 772 3234

**Abstract**

By solving the Young Laplace equation of capillary hydrostatics one can accurately determine equilibrium shapes of droplets on relatively smooth solid surfaces. The solution, however, of the Young Laplace equation becomes tricky when a droplet is sitting on a geometrically patterned surface and multiple, and unknown a priori, three phase contact lines have to be accounted for, since air pockets are trapped beneath the liquid droplet. In this work, we propose an augmented Young-Laplace equation, in which a unified formulation for the liquid/vapor and liquid/solid interfaces is adopted, incorporating microscale interactions. This way, we bypass the implementation of the Young's contact angle boundary condition at each three phase contact line. We demonstrate the method's efficiency by computing equilibrium wetting states of entire droplets sitting on geometrically structured surfaces, and compare the results with those of mesoscopic Lattice Boltzmann simulations. The application of well-established parameter continuation techniques enables the tracing of stable and unstable equilibrium solution families, including the well-known Cassie-Baxter and Wenzel states. The computation of unstable solutions is necessary for the determination of energy barriers separating co-existing stable wetting states. Since energy barriers determine whether a surface facilitates or inhibits certain wetting transitions, their computation is important for many technical applications. Our continuum-level analysis can readily be applied to patterned surfaces with increased and unstructured geometric complexity, having a significant computational advantage, as compared to the computationally demanding mesoscopic simulations that are usually employed for the same task.

## 1. Introduction

The modification of the wetting properties of solid surfaces is important in many contemporary applications (microfluidic devices, controllable drug delivery, self-cleaning surfaces etc.[1]). The wettability can be controlled through thermal [2], chemical [3], magnetic [4] or electric [5] actuation. It is also reported that the range of wettability switching is considerably extended by surface roughness [6]. Artificially roughened surfaces are designed to mimic the morphology of the lotus leaf [7], where in place of the natural surface sculptures, pillared or honeycomb-structures are fabricated. In the case of a droplet sitting on top of these protrusions preserving an almost spherical shape, the surface can be characterized as superhydrophobic [8]. However this state -commonly called the "fakir" or Cassie-Baxter state- is usually metastable, since the droplet

(when properly perturbed) can undergo a collapse transition and impale the solid protrusions to form states of enhanced apparent wettability [9]. This collapsed state -commonly called Wenzel state- features large contact angle hysteresis, due to the strong pinning of the contact line at the pillar corners [10].

Wetting transitions between Cassie-Baxter and Wenzel states occur in an asymmetric fashion. In the case of electrostatic enhancement of material wettability (utilizing the electrowetting phenomenon [11]), collapse transitions occur when certain wettability thresholds are surpassed. However, the reverse transition cannot be realized by switching off the electric actuation. Experimental studies report that strong external actuations (e.g., thermal heating) are required in order to induce de-wetting transitions [12]. From a thermodynamic point of view Cassie-Baxter and Wenzel states correspond to distinct minima of the free energy landscape suggesting the existence of at least one intermediate unstable state. The difference between the saddle of the free energy landscape and the local minima determine the energy barrier for a wetting or a de-wetting transition.

In order to facilitate reversible wetting transitions, proper design of the solid surface roughness is required minimizing the energy barriers, which separate the distinct wetting states [13, 14]. Previous theoretical studies provide analytical expressions for such energy barrier computations, however they are based on significant simplifications regarding the actual shape of the droplet and/or the solid surface [15, 16]. For an accurate determination of the wetting transition energy barriers, the computation of all admissible states, stable and unstable, is required.

The most efficient computational approach for the determination of equilibrium wetting states, is the solution of the Young-Laplace (YL) equation [17]. In this continuum-level method, the solid material wettability is incorporated through the implementation of the Young's contact angle ($\theta_Y$) boundary condition at the three phase contact line (TPL). However, its implementation on rough surfaces is of limited efficiency, since wetting states with multiple three phase contact lines -the number and position of which is unknown- can be admitted.

In particular, YL equation fails to predict wetting states, where air pockets are trapped beneath the liquid droplet (Cassie-Baxter states). The YL formulation is limited to predict the shape of a liquid/air interface beneath the liquid droplet, by performing computations in a unit cell [18].

Cassie-Baxter states are well predicted by implementing appropriately initialized dynamic simulators, such as molecular dynamics [19, 20] and mesoscopic Lattice-Boltzmann (LB) models [21, 22]. These models take into account intermolecular interactions between liquid/vapor, liquid/solid and solid/vapor phases. However, long time executions are required in order to converge to an equilibrium state. Moreover, such equilibrium states - obtained through direct dynamic simulations - can only be stable, excluding the computation of energy barriers. A time-stepper based methodology that enables the computation of both stable and unstable solutions has been proposed in [21], still the computational cost may be prohibitive for real scale roughness simulations. Alternatively, and in order to preserve the computational cost at a reasonable level, the continuum-level methods can be reformulated so as to encompass molecular details, lumped into a solid/liquid interaction parameter.

An attempt to incorporate microscale liquid/solid (LS) interactions in the YL equation can be found in [23, 24], for droplets wetting flat solid surfaces and in [25] for the case of geometrically structured surfaces; however, the computation of Cassie-Baxter wetting states is not feasible, due to the assumption that a thin liquid film always wets the solid surface. In this work, we propose a new formulation of the YL equation augmented with a term modeling a disjoining pressure. This augmented YL equation is parameterized in terms of the arc-length of the effectively one dimensional droplet surface, and enables the treatment of the liquid/vapor and the liquid/solid interface in a unified context. By adopting this formulation, we track the entire droplet surface which is in contact either with the vapor phase (conventional YL formulation), or the solid material. This way, droplet profile computations with multiple three phase contact lines can be performed, unencumbered by the need to implement the Young's contact angle boundary condition. In this formulation, the Young's contact angle emerges "naturally" as the result of the combined action of the disjoining pressure (active at the liquid/solid interface [26]) and the surface tension (liquid/vapor interface). Finally, by applying pseudo arc-length parametric continuation [27], we can detect stable and unstable wetting states, that are necessary for energy barrier computations with significantly lower computational requirements compared to time-stepper based approaches which utilize LB simulations.

This article is organized as follows: We first present the mathematical formulation of the augmented YL equation for cylindrical liquid droplets. The augmented YL is validated with results obtained from the conventional YL equation, for droplets wetting flat and single-striped solid surfaces. Its implementation on multi-striped solid surfaces is compared with the LB method described in [21]. Finally, we explore the entire solution space of a droplet wetting a striped-patterned solid surface and present accurate computations of wetting and de-wetting energy barriers.

## 2. Mathematical formulation

2.1. Augmented Young-Laplace equation

The YL equation states the balance of mechanical forces along the liquid/vapor interface. When augmented with a disjoining pressure term, modeling the liquid/solid interactions, it reads:

$$\frac{\gamma_{LV}}{R_o} C + p^{LS} = \Delta p \Rightarrow$$

$$C + \frac{R_o}{\gamma_{LV}} p^{LS} = K, \tag{1}$$

where $\gamma_{LV}$, is the liquid/vapor (LV) interfacial tension, $\Delta p$, is the pressure difference between the liquid and the surrounding vapor phase. $K \equiv R_o \Delta p / \gamma_{LV}$ is a reference dimensionless pressure, which is constant along the droplet surface and $R_o$, is a characteristic length (droplet radius). $C$ is the dimensionless local mean curvature of the droplet surface. The disjoining pressure, $p^{LS}$, incorporates the van der Waals interactions, as well as the electrostatic interactions attributed to overlapping of electrical double layers of liquid/solid interfaces [28, 29]. In order to model i) strong repulsion exerted to the liquid phase when being at close proximity, and ii) attraction at an intermediate Euclidean distance, $\delta$, from the solid material, we adopt the following formulation for $p^{LS}$ (resembling a Lennard-Jones potential) in dimensionless form:

$$\frac{R_o}{\gamma_{LV}} p^{LS}(\delta) = w^{LS} \left[ \left(\frac{\sigma}{\delta + \varepsilon}\right)^{C1} - \left(\frac{\sigma}{\delta + \varepsilon}\right)^{C2} \right], \tag{2}$$

where $w^{LS}$ is a dimensionless wetting parameter, which controls the wettability of the solid by the liquid; increasing $w^{LS}$ values enhances solid/liquid affinity. The $C1$, $C2$ constants (with $C1 > C2$) control the range of action of the microscale forces; $\sigma$ regulates the minimum distance between the liquid and the solid phase. When the solid surface is flat, $\delta$ is the dimensionless vertical distance of the droplet surface from the solid. For non-flat (geometrically patterned) surfaces, $\delta$ cannot be defined in a straightforward manner. Here, for the first time we propose, $\delta$, to be defined as the solution of the Eikonal equation [30] which determines the signed distance from a solid boundary. The added computational cost is not significant, since a single Eikonal equation solution is required for fixed particular solid surface structure. A discussion of the numerical solution of the Eikonal equation is provided in the appendix (A1). Finally, $\varepsilon$, is a small constant number used in order to avoid numerical singularities when $\delta \to 0$.

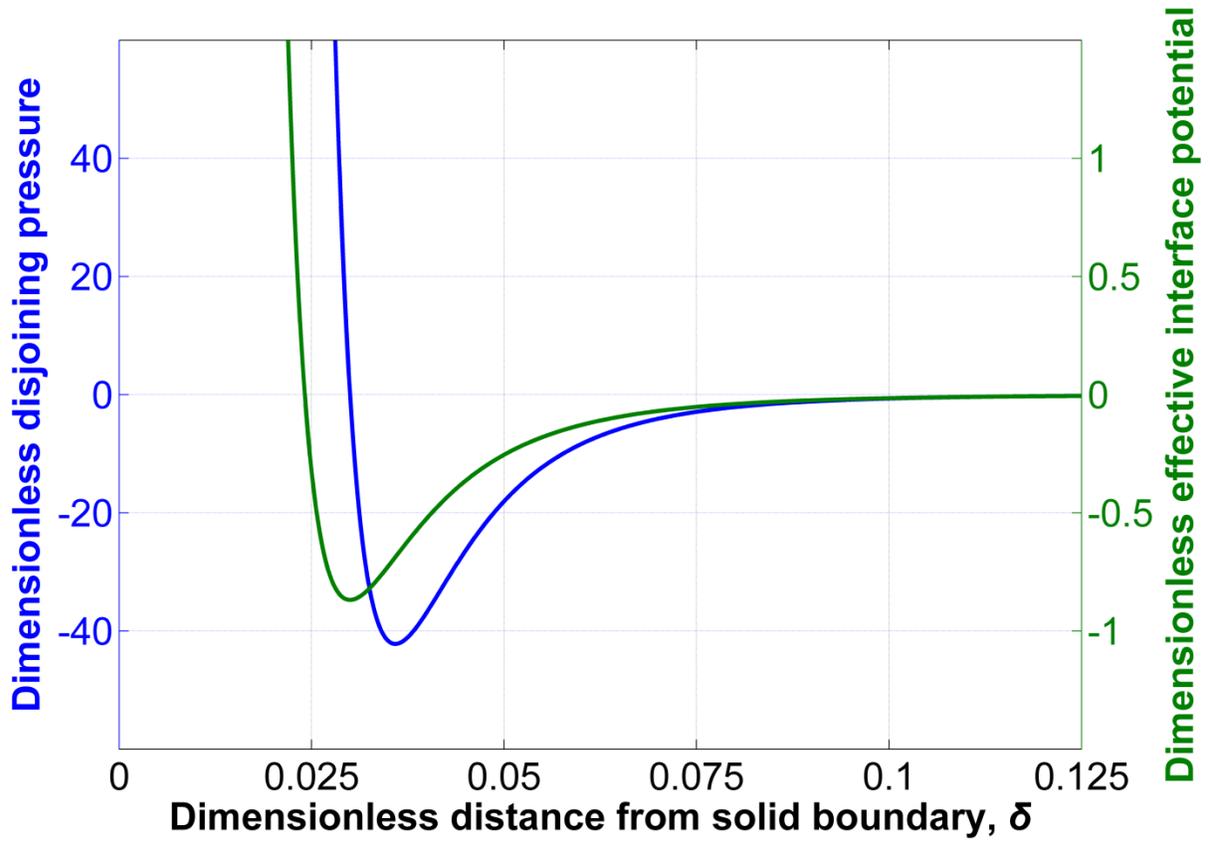

Fig. 1 – Dependence of the dimensionless disjoining pressure $p^{LS} R_o/\gamma_{LV}$ and of the corresponding dimensionless effective interface potential $\omega/\gamma_{LV}$ from the distance $\delta$.
$\sigma = 3.8\text{x}10^{-2}$, $C1 = 8$, $C2 = 6$, $\varepsilon = 8\text{x}10^{-3}$, $w^{LS} = 4\text{x}10^{2}$.

The disjoining pressure profile we employ in this work (see Fig. 1) models a non-wetting isotherm [28, 31]. Different modeling approaches can also be adopted for the inclusion, e.g. of liquid adsorption on the solid surface.

2.2. Arc-length parameterization of the Young-Laplace equation

The proposed formulation is demonstrated for cylindrical droplets wetting textured surfaces (see Fig. 2) assuming translational symmetry along the direction perpendicular to the *xy*-plane. The droplet surface is conveniently defined in cylindrical coordinates $(r, \varphi)$. By parameterizing the droplet surface with the angular coordinate, $\varphi$, (i.e., $r \equiv r(\varphi)$) the problem becomes one-dimensional. This parameterization is sufficient for smooth or single-striped solid surfaces, where $r(\varphi)$ is a single value function. When wetting occurs on solid surfaces of increased topographic complexity different $r$ values can correspond to the same angular coordinate, thus the angular parameterization is deficient. An alternative approach is to parameterize the droplet surface as a function of its arc-length, $s$ (see Fig. 2), (i.e., $r \equiv r(s)$, $\varphi \equiv \varphi(s)$). This parameterization enables the tracking of the droplet surface in a natural way, incorporating the liquid/vapor and liquid/solid interactions, while the problem is maintained one-dimensional.

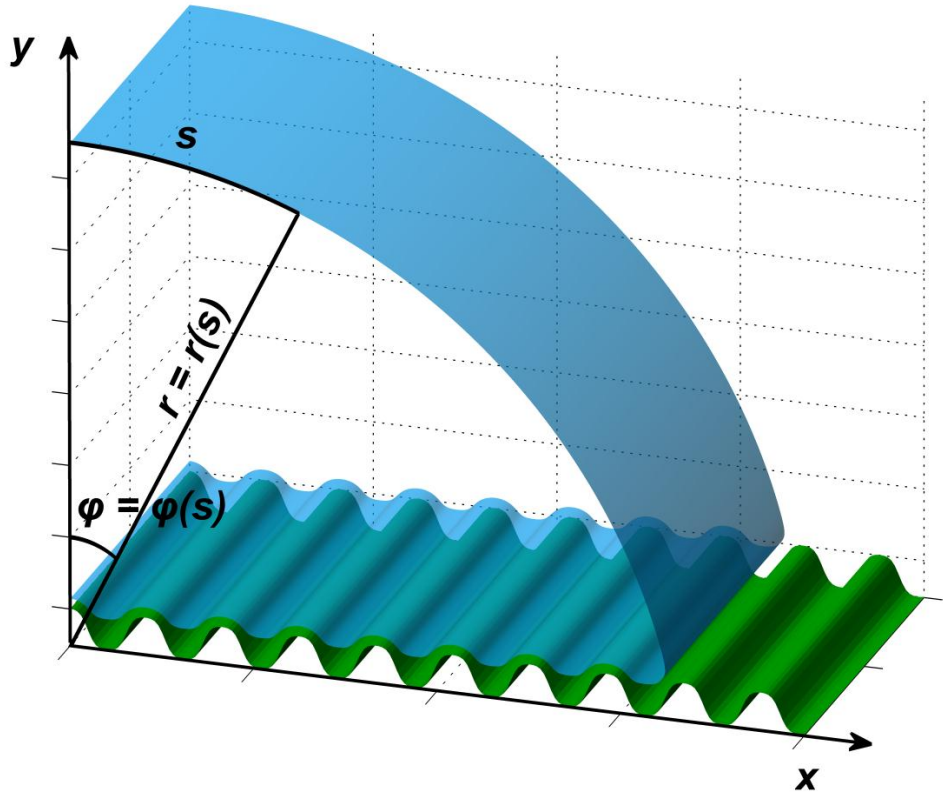

Fig. 2 – Cylindrical liquid droplet profile on a stripe-patterned surface.

The local mean curvature of the droplet's surface formulated as a function of *s*, reads:

$$C = \frac{1}{\sqrt{r^2\varphi_s^2 + r_s^2}}\left[\varphi_s + \frac{\partial}{\partial s}\arctan\left(\frac{r\varphi_s}{r_s}\right)\right], \tag{3}$$

$$\varphi_s \equiv \frac{\partial \varphi}{\partial s}, \quad r_s \equiv \frac{\partial r}{\partial s}.$$

The unknown functions, *r(s)* and *φ(s)*, are determined by the solution of the augmented Young-Laplace equation and the dimensionless arc-length differential equation, which defines the incremental arc-length of the droplet surface:

$$dr^2 + r^2 d\varphi^2 = ds^2 \Rightarrow$$

$$\left(\frac{dr}{ds}\right)^2 + r^2\left(\frac{d\varphi}{ds}\right)^2 = 1 \tag{4}$$

The incompressibility of the liquid is expressed by the following dimensionless integral constraint for the conservation of the droplet's cross sectional area; when $A_{\text{droplet}} = \pi$:

$$\int_0^{s_{max}} r^2 \varphi_s ds = \pi. \tag{5}$$

The maximum arc-length of the droplet surface $s_{max}$, which defines the boundary of the computational domain, is unknown. This suggests a free boundary problem. To deal with this extra unknown we introduce the following algebraic equation:

$$\varphi = 0 \text{ at } s = s_{max}. \tag{6}$$

Moreover, Neumann-type boundary conditions are imposed both at the apex and at the bottom of the droplet, since reflection symmetry is considered about the vertical plane ($\varphi = 0$).

$$\frac{\partial r}{\partial s} = 0 \text{ at } s = 0 \text{ and at } s = s_{max}. \tag{7}$$

Finally, the following Dirichlet-type boundary condition singles out the arc-length equation solution and it is imposed at the apex of the droplet:

$$\varphi = 0 \text{ at } s = 0. \tag{8}$$

The differential equations (1) and (4) with the integral equation (5) and the algebraic equation (6) are discretized using the Galerkin finite element method [32] and the resulting set of non-linear equations, $R(u \equiv (r,\varphi,K,s_{max})^T)=0$, is solved iteratively (with Newton-Raphson) for the droplet surface coordinates $r(s)$ and $\varphi(s)$, the maximum arc-length, $s_{max}$, and the reference pressure, $K$. The solution space is explored through parameter continuation methods, and in particular the pseudo arc-length continuation technique [27], which enables the computation of stable and unstable steady state solutions. The stability of the computed solutions is quantified solving the eigenvalue problem of the Jacobian matrix, $\partial R/\partial u$. One solution is stable when the eigenvalues of the Jacobian matrix are negative; when at least one eigenvalue is positive, the solution is characterized as unstable [33].

2.3. Energy barrier computations

The computation of wetting or de-wetting energy barriers implies the existence of stable states multiplicity for a certain material wettability i.e. for a certain value of the wetting parameter. Such stable wetting states correspond to distinct minima of the surface free energy; this picture suggests the existence of at least one intermediate local maximum of the free energy corresponding to an unstable state. This unstable state, which is not observed in experiments, corresponds to a saddle of the surface free energy landscape. The difference between the saddle and the local minima of energy determines the energy barrier for the realization of wetting (or de-wetting) transitions. In small liquid droplets (<10 μL), the effect of gravity can be neglected, thus its surface energy can be expressed as:

$$F = \gamma_{LV} A_{LV} + \gamma_{LS} A_{LS} + \gamma_{SV} A_{SV}, \tag{9}$$

where $\gamma_{LS}$ and $\gamma_{SV}$ are the interfacial tensions of the liquid/solid (LS) and the solid/vapor (SV) interfaces and $A_{LV}$, $A_{LS}$ and $A_{SV}$ are the surface areas of (LV), (LS) and (SV) interfaces respectively. Given the Young's equation [34]:

$$\gamma_{LV} \cos\theta_Y + \gamma_{LS} = \gamma_{SV}, \tag{10}$$

the surface free energy reads:

$$F = \gamma_{LV}(A_{LV} - \cos\theta_Y A_{LS}) + \gamma_{SV} A_{solid}, \tag{11}$$

where $A_{solid} = A_{LS} + A_{SV}$ is the total solid surface area. If we denote with $F_{su}$, $F_{co}$ and $F_{un}$ the surface free energy of the stable suspended (*su*), stable collapsed (*co*) and unstable (*un*) - intermediate wetting states (Fig. 3), which can be admitted on a textured solid surface with a constant Young's contact angle, $\theta_Y$, then the wetting (or the suspended to collapsed, $su \to co$) energy barrier can be computed from:

$$E_{su \to co} = F_{un} - F_{su} \Rightarrow$$

$$\frac{E_{su \to co}}{\gamma_{LV}} = (A_{LV,un} - A_{LV,su}) - \cos\theta_Y (A_{LS,un} - A_{LS,su}). \tag{12}$$

Thus, energy barrier computation requires the calculation of the surface area of the LV and LS interfaces of the unstable and stable wetting steady states for a given Young's contact angle value, $\theta_Y$.

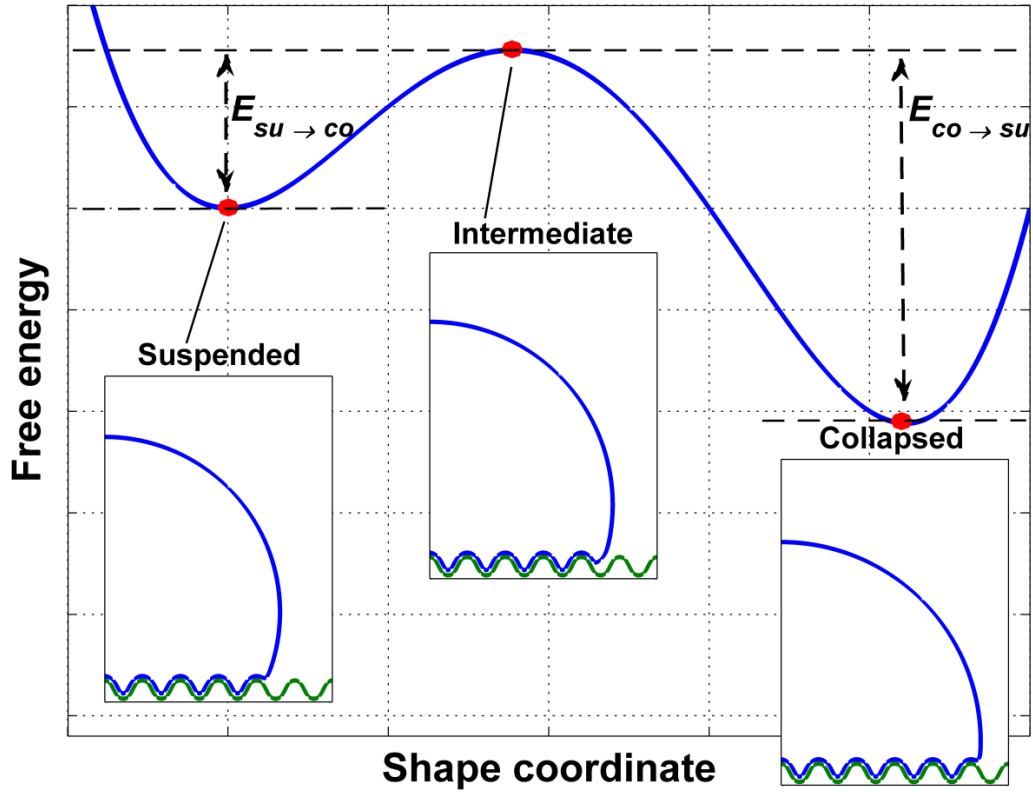

Fig. 3 – Schematic of a segment of the surface free energy profile for a textured solid surface and constant Young angle, $\theta_Y$.

The surface area of these interfaces can be calculated as: $A = S\, L$, where $S$ is the corresponding arc-length of the interface and $L$ expresses the depth along the direction perpendicular to the *xy*-plane; because of the translational symmetry of the droplet, $L = 1$. The energy barrier for a de-wetting transition, $E_{co \to su}$, is computed similarly.

2.4. Wetting parameter and Young's contact angle correlation

As reported above, contrary to the conventional YL equation, in which the Young's contact angle, $\theta_Y$, is introduced explicitly as a boundary condition, in the augmented YL, the material wettability is accounted for through the wetting parameter value, $w^{LS}$. However, the Young's contact angle estimation is required for the energy barriers computations. Therefore, a correlation between this wetting parameter and the Young's contact angle should be derived as follows:

The effective interface potential, $\omega$, (see Fig. 1) which represents the cost in free energy per unit area to maintain a distance $\delta$ between the solid and the liquid surfaces (with $\omega \to 0$, when $\delta \to \infty$) is related with the disjoining pressure $p^{LS}$, according to [35]:

$$p^{LS} = -\frac{d\omega}{d\delta}. \tag{13}$$

The effective interface potential reaches its minimum value, $\omega_{min}$, at the liquid/solid interface, and the following relation holds [35, 36]:

$$\omega_{min} = \gamma_{SV} + \gamma_{LV} - \gamma_{LS}. \tag{14}$$

Finally, using the Young's equation yields:

$$\cos\theta_Y = \frac{\omega_{min}}{\gamma_{LV}} - 1, \tag{15}$$

which correlates the liquid/solid affinity strength ($w^{LS}$) with the macroscopically observed Young's contact angle.

3. **Results and discussion**

3.1. Flat solid surface

We first present computational predictions of the proposed augmented YL equation for the equilibrium of sessile droplets on smooth and flat solid surfaces with variable wettability i.e., by varying the wetting parameter. To validate our findings we determine the Young's contact angle value by performing circular fitting to the liquid/vapor interface (see appendix A2) and compare with eq. (15). The comparison presented in Fig. 4 shows excellent agreement, except for extremely low wetting parameter values ($w^{LS} \to 0$), which do not correspond to realistic systems.

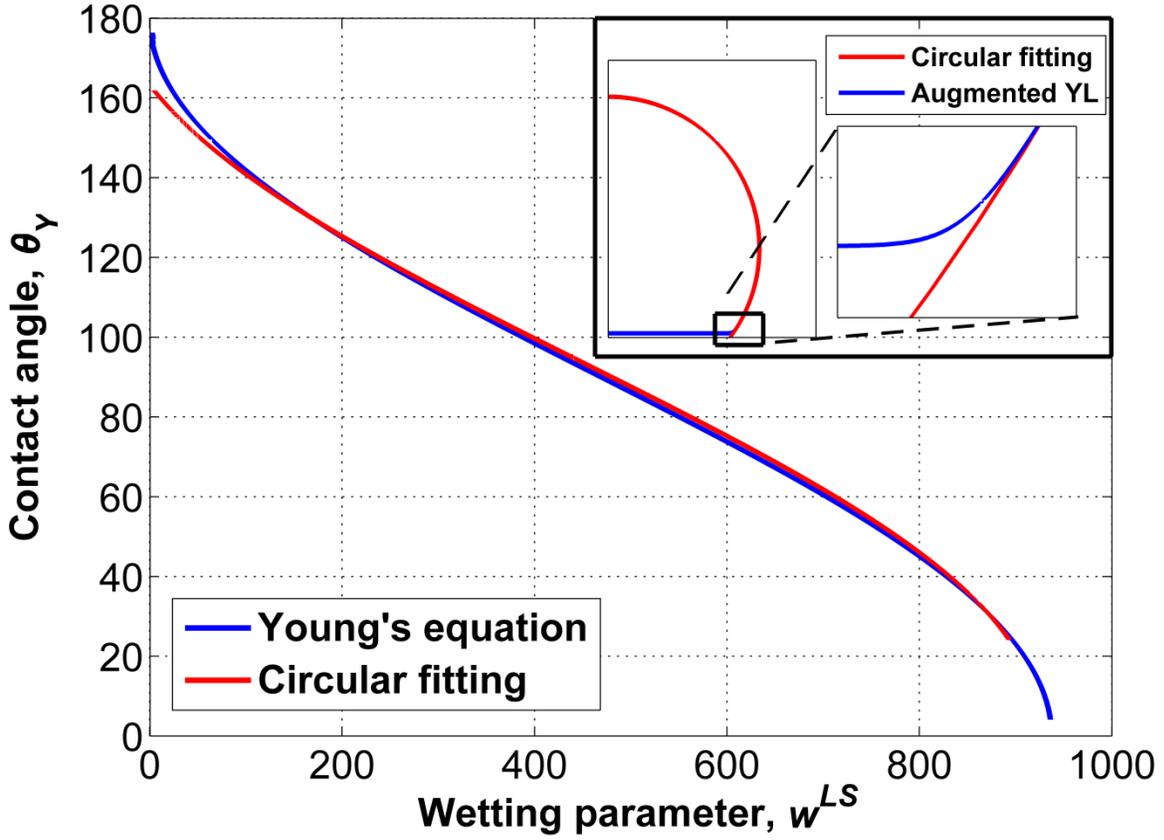

Fig. 4 – Young's contact angle, $\theta_Y$ for droplet on a flat surface as a function of the $w^{LS}$. The blue line is obtained from the application of Young's equation (15); the red line is produced by circular fitting to the liquid/vapor interface – Embedded figure: Circular fitting to the liquid/vapor interface obtained by the augmented YL equation.

In Fig. 5 we illustrate the variation of the local mean curvature along the surface of the droplet, when $\theta_Y = 120°$. One can observe three interfacial regions: (I) The liquid/vapor interface region, where the curvature is constant, (II) the region in the vicinity of the TPL, with sharply increasing curvature, and (III), the liquid/solid interface region where the curvature is equal with the curvature of the solid surface (zero for flat surfaces). The conventional YL equation treats only the region (I) up to the TPL, where the Young's boundary condition is applied. The augmented YL equation accounts for the entire droplet surface, including regions where liquid/solid interactions are active ((II) and (III)).

The electrostatic interactions of liquid/solid interfaces [28, 29] induce a minimum distance ($\delta_{min}$) between the droplet and the solid boundary. When $\delta = \delta_{min}$, then the disjoining pressure, $p^{LS} = 0$. For $C1 = 8$ and $C2 = 6$, eq. (2) yields:

$$\delta_{min} = \sigma - \varepsilon. \tag{16}$$

The minimum $p^{LS}$ value corresponds to the three phase contact line at which the local mean curvature of the droplet is maximum. The distance of the liquid from the solid surface at the TPL, $\delta_{TPL}$, reads:

$$\delta_{TPL} = \frac{2\sqrt{3}\sigma}{3} - \varepsilon. \tag{17}$$

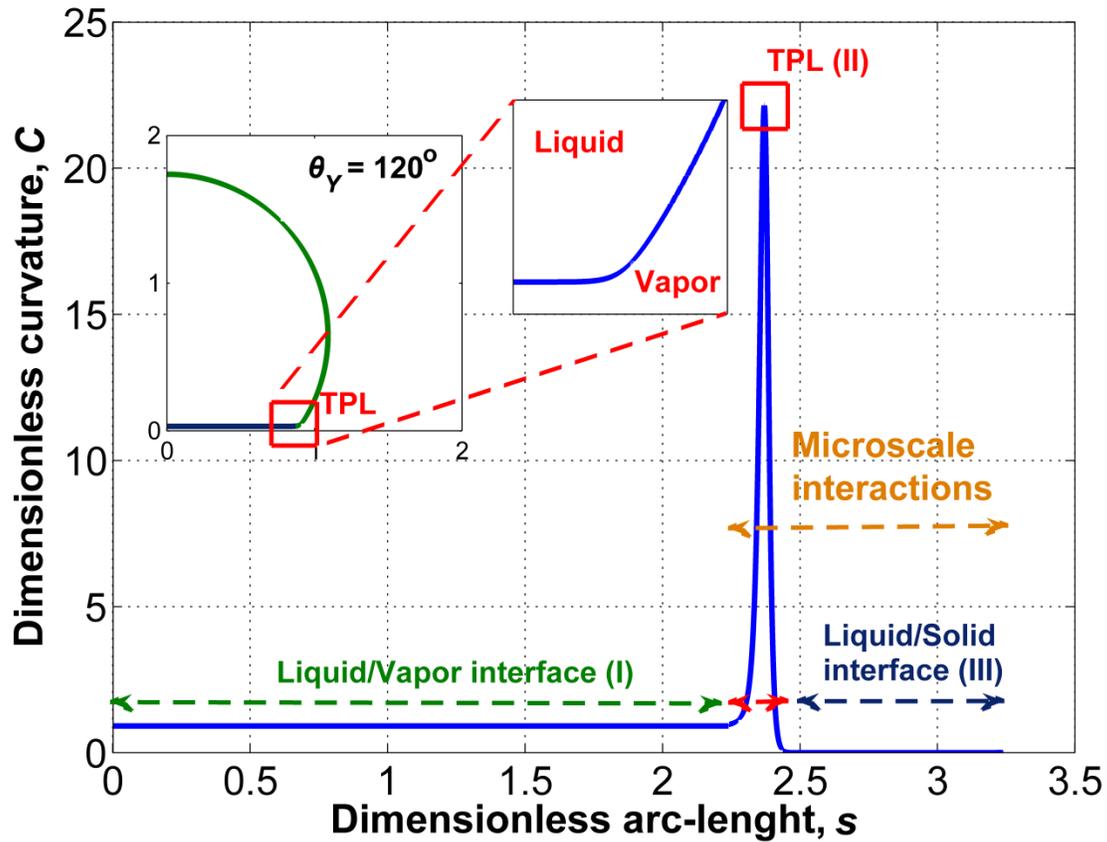

Fig. 5 – Dimensionless curvature ($C$) variation along the arc-length for a droplet wetting a flat solid surface with $\theta_Y = 120°$.

3.2. Single-stripe structured surface

In this section we test the validity of the augmented YL equation by comparing its predictions with the ones obtained from the conventional YL equation for a single-stripe structured solid surface. This is a quite simple solid surface geometry that can sustain both stable and unstable solutions which co-exist and can also be treated with the conventional YL equation as shown in [21]. The stripe shape is given by:

$$y = \frac{1 - \mathrm{erf}((x-p_1)p_2)}{p_3}, \tag{18}$$

where erf denotes the error function. The parameters $p_1$, $p_2$ and $p_3$ determine the geometric features of the stripe: $p_1$ regulates the width of the stripe at its top surface, $p_2$ controls the curvature of the lateral walls and $p_3$ determines the maximum height.

The bifurcation diagram in Fig. 6 shows the dependence of the dimensionless droplet height on the material wettability ($\theta_Y$). For the augmented YL equation the bifurcation parameter is the wetting parameter, $w^{LS}$, which can be correlated with the Young's contact angle, $\theta_Y$, through eq. (15).

One can observe that both formulations predict, with exceptional agreement, an S-shaped curve with unstable wetting states (dashed lines) linking the branches of stable wetting states (solid lines) at the two turning points

((A) and (C)). This bifurcation diagram shows that transitions – induced by wettability modification – between suspended (upper stable branch) and collapsed states (lower stable branch) are hysteretic. Starting with a hydrophobic material (point (D)) and gradually enhancing material wettability (e.g., through electrowetting), then a drop collapse occurs at turning point (A) ($\theta_Y \approx 79°$) i.e., the drop collapses following the route (D) → (A) → (B).

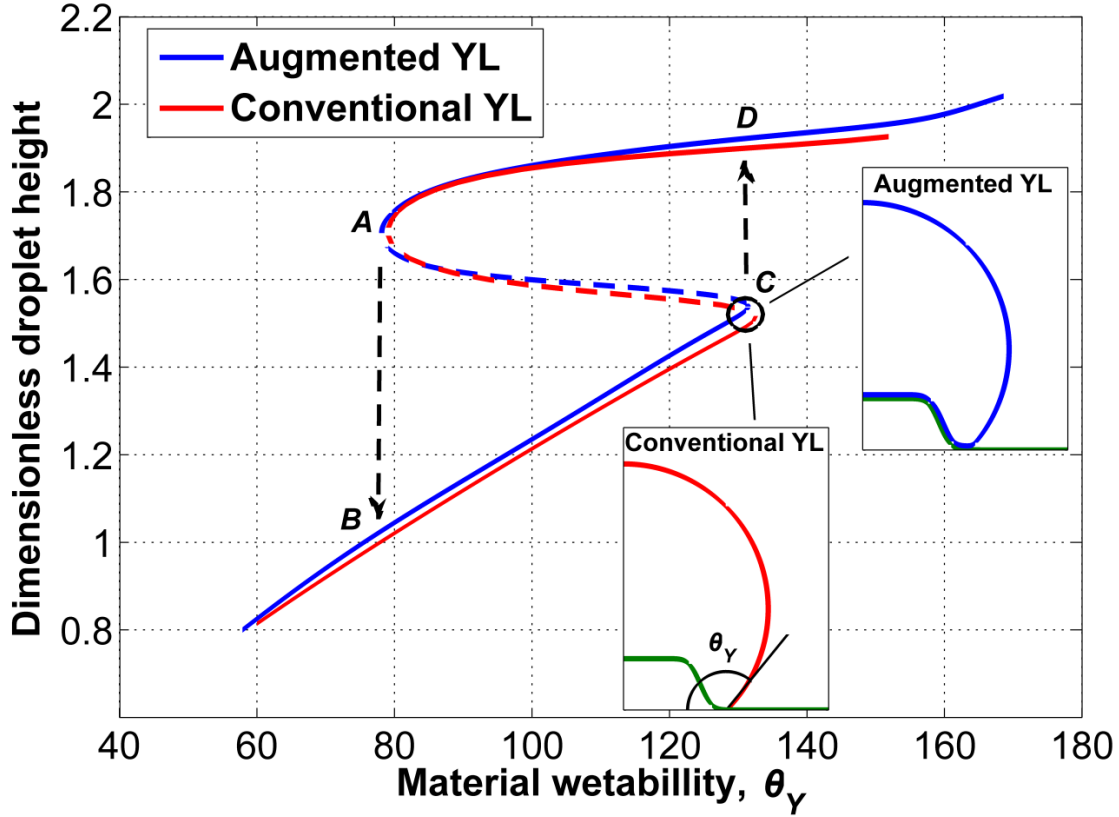

Fig. 6 – Dependence of the droplet height on the Young contact angle ($\theta_Y$) for a single-striped solid surface structure ($p_1 = 0.6$, $p_2 = 10$, $p_3 = 5$).

In a reverse experiment, where wettability is decreased, the droplet undergoes a de-wetting transition at turning point (C). For this particular geometry, this lifting transition would require lowering of wettability of as much as $\theta_Y \approx 130°$. This suggests that de-wetting through wettability modification cannot be realized, considering that common hydrophobic materials do not exhibit $\theta_Y > 130°$ (e.g. for Teflon® AF 1600, $\theta_Y = 120°$). In experimental practice the range of tunable wettability is limited; e.g., when wettability is electrostatically enhanced, $\theta_Y$ has an upper limit set by the material chemistry [37] and a lower one due to the contact angle saturation phenomenon [38].

3.3. Patterned surface – Comparison with mesoscopic LB simulations

The augmented YL formulation can be easily applied to the equilibrium of a droplet sitting on a structured surface of increased topographic complexity, e.g. a stripe-patterned surface. The predictions of the proposed methodology are compared with the ones obtained from LB simulations. A detailed description of the employed LB model can be found in [39]. The LB model accounts for non-trivial microscopic effects (e.g.

disjoining pressure, partial wetting) through the application of fluid-solid pseudo-potential interactions. The solid-liquid interactions include an attractive part modeling London-van der Waals forces, and a repulsive component associated with double-layer interactions and in direct analogy with DLVO theory [28, 29]. Fluid-fluid interactions are defined to ensure constant compressibility in both phases, with a density ratio 10:1 and a compressibility ratio 15.625:1 when the system reaches phase equilibrium.

The solid surface is structured with wave-like stripes the height of which is described by:

$$y = \left(\frac{\cos(3x)}{1.8}\right)^2. \tag{19}$$

The dependence of the apparent contact angle ($\theta_a$) as a function of the material wettability, $\theta_Y$, is depicted in Fig. 7 for both approaches: the continuum-level augmented YL and the mesoscopic LB model. The apparent contact angle is calculated by fitting a circle to the liquid/vapor interface above the solid surface asperities. The droplet interface obtained from the LB application is defined by the density contour corresponding to $\rho = (\rho_{Liquid}+\rho_{Vapor})/2$, where $\rho_{Liquid}$, $\rho_{Vapor}$ are the densities of the liquid and vapor bulk phase, respectively.

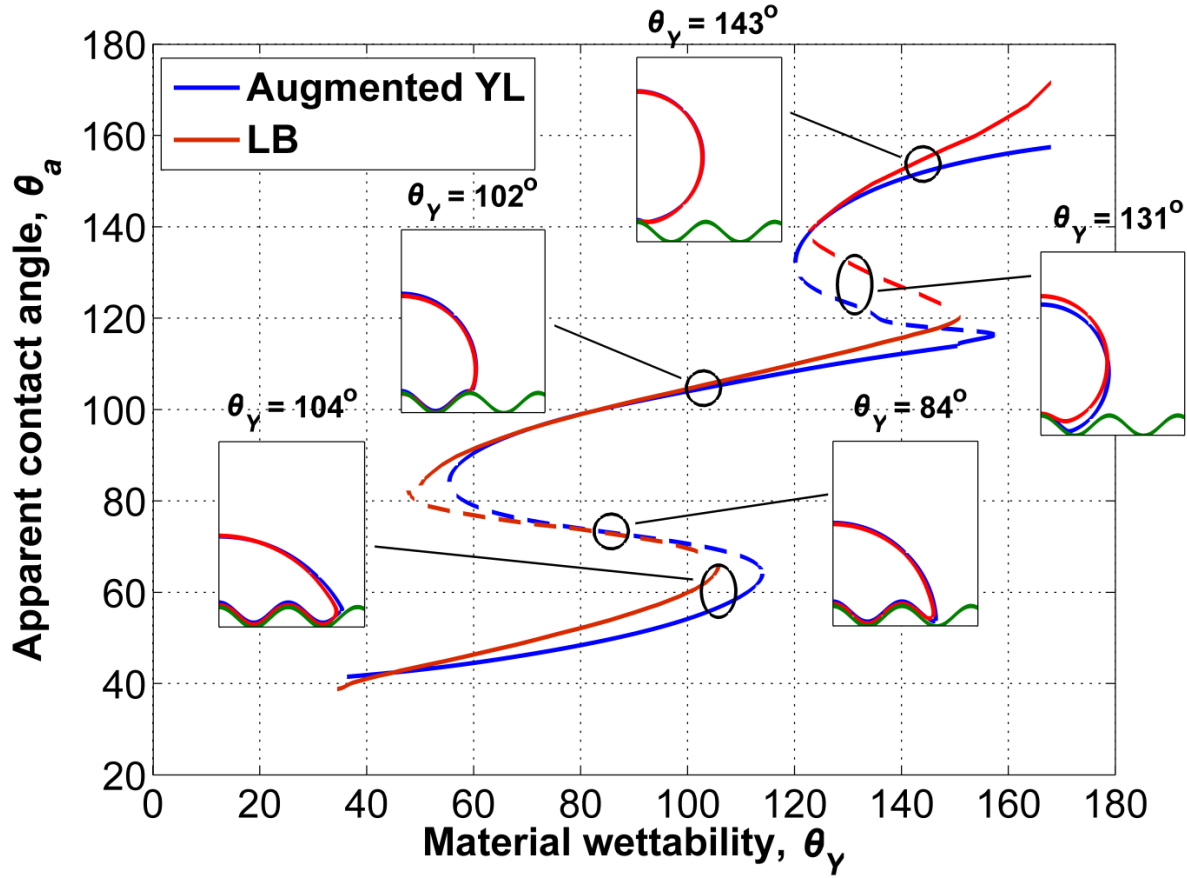

Fig. 7 – Dependence of the apparent contact angle ($\theta_a$) on the material wettability ($\theta_Y$) for a patterned solid surface. The blue line corresponds to the results obtained from the augmented YL equation. The red line corresponds to LB – based computations.

Despite the fact that the YL and the LB are conceptually different descriptions of wetting phenomena (YL: continuum-level, LB: mesoscopic), they both capture, with satisfactory, agreement the non-trivial macroscopic behavior of a liquid droplet wetting a striped structure. The observed discrepancies are attributed to the diffuse

LS and LV interfaces modeled in the LB description, as well as the different formulation of the liquid/solid interactions.

We re-iterate that the augmented YL description is a one-dimensional problem, as opposed to the two dimensional LB formulation. The computational benefits of the proposed continuum-level analysis are significant; indicatively we report that a Newton-Raphson iteration utilizing LB runs requires approximately 10 minutes, whereas the augmented YL model requires about 1 second, when all computations are performed on the same CPU (Intel® Core™ i7 @3.07 GHz). This computational advantage of the augmented YL equation is obtained without implementing any adaptive mesh technique for automatic refinement in regions of high interface curvature. Consequently, the use of the augmented YL equation is a very efficient and accurate method to model wetting phenomena on structured surfaces, incorporating microscale interactions.

### 3.4. Patterned surface of increased roughness

#### 3.4.1. Parametric analysis

The proposed formulation is used to compute the admissible equilibrium wetting states on a stripe-patterned solid surface of increased roughness. The topography of the surface is given by the following relation:

$$y = 1 - \frac{erf(1.3\cos(33x+3.3))}{20}. \tag{20}$$

The surface roughness (ratio of the actual over the apparent surface area) is 1.52 as opposed to 1.19 for the solid surface presented in the previous section. The unit cell of this solid surface structure is a stripe (see eq. 18) with $p_1 = 4.31 \times 10^{-2}$, $p_2 = 43.88$, $p_3 = 21.16$. The dimensionless wavelength period is $9 \times 10^{-2}$. It is worth mentioning that the application of bifurcation analysis utilizing LB simulations, for this case study, requires substantial amount of computational time and resources. However, the solution of the augmented YL equation sustains its computational needs at very low levels.

In Fig. 8 we present the dependence of the apparent contact angle ($\theta_a$) on the material wettability, quantified by $\theta_Y$. The diagram exhibits multiple S-shaped curves, which are directly associated with the number of stripes of the solid surface. Multiple wetting steady states co-exist within large $\theta_Y$ value intervals, e.g. 9 wetting states co-exist for a solid material with $\theta_Y = 104°$. Wetting or de-wetting transitions induced by material wettability modification set on at the turning point values of $\theta_Y$. One can observe that de-wetting transitions for this particular solid surface would demand a material wettability of $\theta_Y > 120°$ ($\theta_Y$ value at the lower right side turning point) which is not feasible with the common hydrophobic materials.

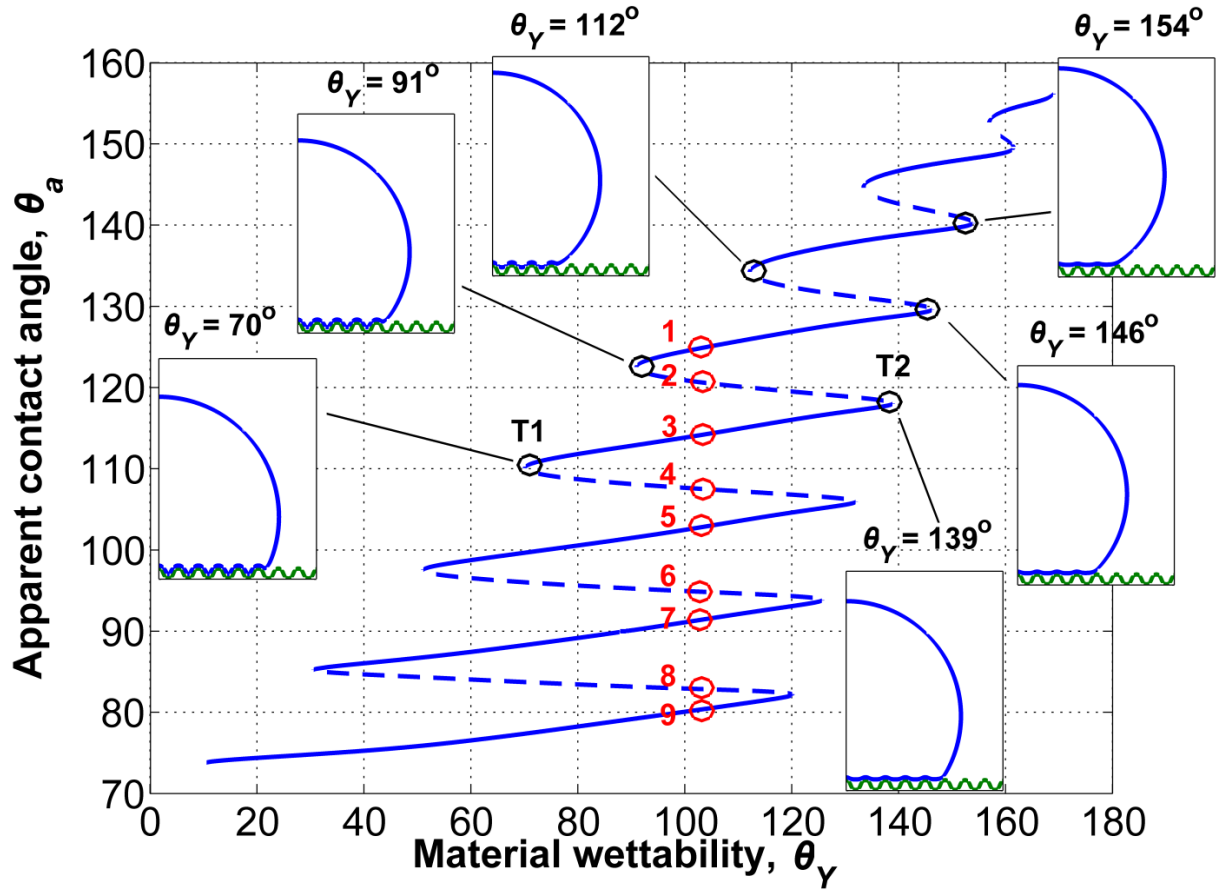

Fig. 8 – Dependence of the apparent contact angle ($\theta_a$) on the material wettability ($\theta_Y$) for a patterned solid surface.

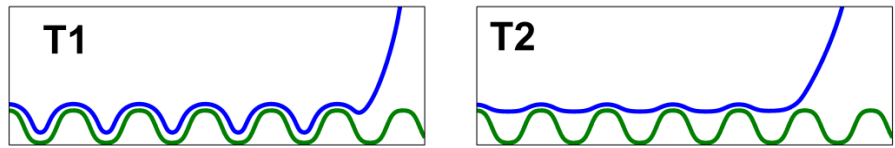

Fig. 9 – Local magnification of the droplet shapes corresponding to points (T1) and (T2) of Fig. 8.

Each stable branch (solid line) extends over a finite interval of $\theta_Y$ values, the maximum of which corresponds to droplets suspended on the surface protrusions trapping air beneath it (see local magnification of the droplet shape corresponding to state (T2) at Fig. 9); the lower end of this interval corresponds to wetting states being immersed in the asperities of the surface (see local magnification of the droplet shape corresponding to state (T1) at Fig. 9) retaining the number of stripes wetted. It is observed that the material wettability modification, along a stable branch, does not result in significant change of the apparent wettability. Indicatively, the difference in the apparent contact angle ($\Delta\theta_a$) between states (T2) and (T1) is $9^o$, whereas the material wettability modification required is far greater ($\Delta\theta_{Y\,(T2\rightarrow T1)} = 69^o$). This change, along the same stable branch, is not usually observed in experiments because the apparent contact angle variation is small.

On the other hand, transitions between co-existing states, wetting a different number of stripes (e.g. from point (3) to (7) at Fig. 8) lead to substantial apparent contact angle change ($\Delta\theta_{a\,(3\rightarrow 7)} = 23^o$) without modifying the wettability properties ($\Delta\theta_{Y\,(3\rightarrow 7)} = 0^o$). Thereafter, the experimentally observed significant apparent wettability modification (e.g. at a Cassie-Baxter to Wenzel transition [15]) is attributed to the variation in the number of stripes covered by the liquid, which can be triggered, e.g., by mechanical actuation [40].

In Fig. 10 we illustrate nine co-existing droplet profiles (five stable and four unstable) for a hydrophobic material with $\theta_Y = 104^o$ covering a large interval of apparent wetting behavior types, i.e. from hydrophilic to super-hydrophobic. In particular, the apparent contact angle, $\theta_a$, of the droplet ranges from $80^o$ up to $125^o$. In experimental practice, a super-hydrophobic state is attained when the droplet is gently deposited on the protrusions of the surface, whereas impaled states require forced droplet deposition (droplet impacts the surface).

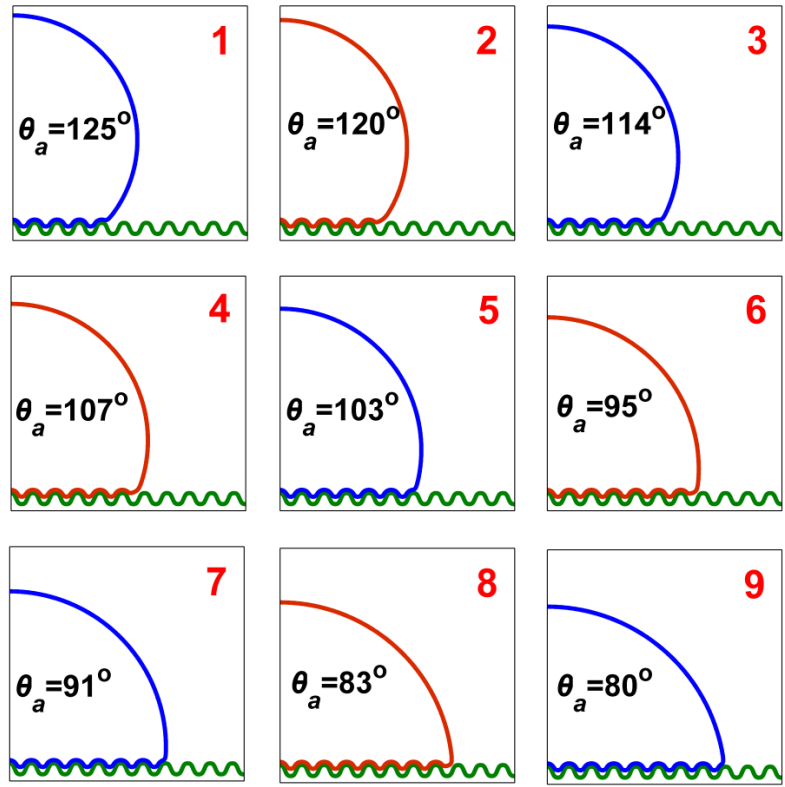

Fig. 10 – Droplet profiles corresponding to points (1) - (9) in figure 8. Profiles drawn with blue line correspond to stable wetting states whereas those with red line correspond to unstable ones. All states computed correspond to material wettability, $\theta_Y = 104^o$.

A wetting transition, e.g. between states (1) and (3) (see also Fig. 8) requires a minimum amount of energy (energy barrier), set by the intermediate unstable state (2). Such transitions that lead to an increased (or decreased) number of stripes cannot be predicted by computational approaches that consider wetting simulations on a unit cell of the structured surface [41]. Such computations disregard the pinning/ de-pining effect of the droplet contact line at the edge of stripes, when more stripes are wetted, and are of acceptable accuracy only when the droplet size is considerably larger than the roughness scale. Nevertheless, accurate energy barrier computations, independently from the roughness scale, can only be performed accounting for the interfacial energies along the surface of the entire droplet. Such computations are presented below.

3.4.2. Energy Barriers Computations

In Fig. 11 we present the energy barriers (using eq. (12)) for wetting and de-wetting transitions on a hydrophobic material with $\theta_Y = 104°$. The admissible stable states, between which transitions can occur, are shown in Fig. 8: (1) → (3) → (5) → (7) → (9) for wetting, and (9) → (7) → (5) → (3) → (1) for de-wetting. It is found that the energy barriers increase as the number of the wetted stripes increases. Notice that $E_{(1)\to(3)} < E_{(3)\to(5)} < E_{(5)\to(7)} < E_{(7)\to(9)}$. On the contrary, for de-wetting transitions the energy barriers increase as the number of the wetted stripes decreases.

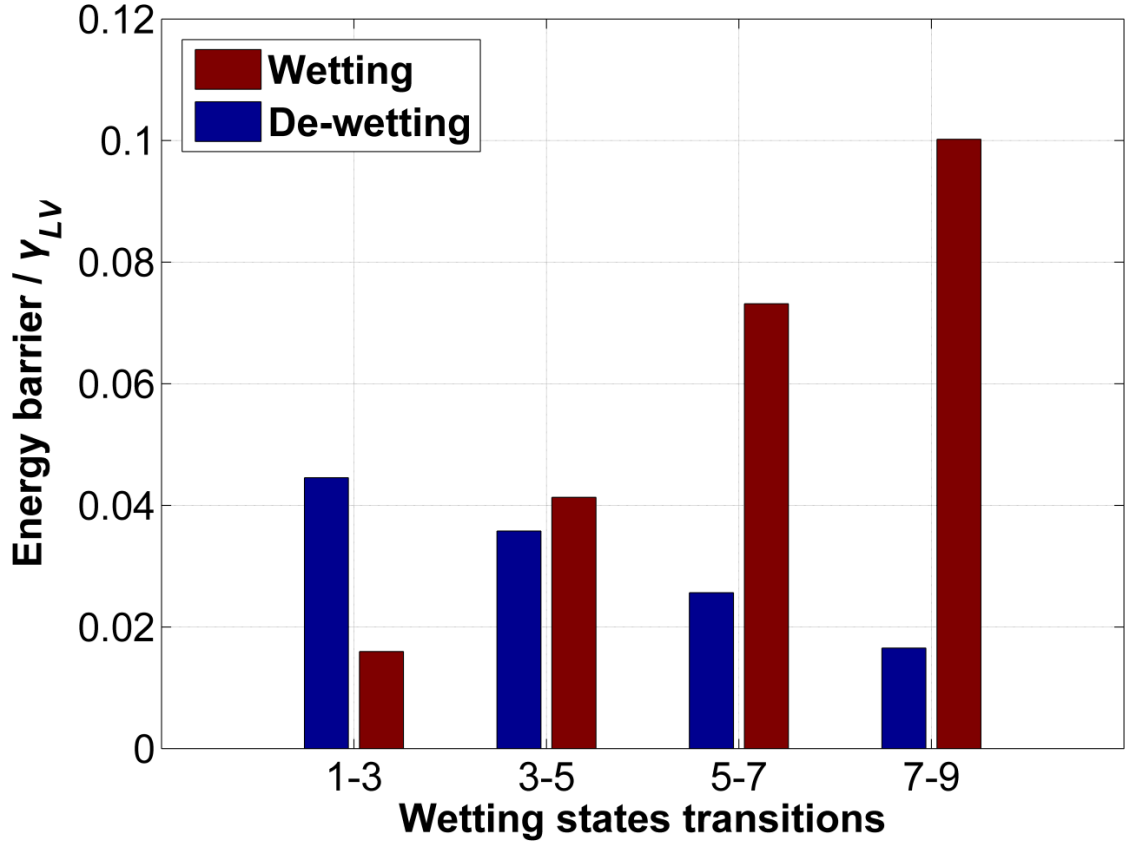

Fig. 11 – Energy barriers computed for the wetting and de-wetting transitions between the stable wetting states corresponding to points (1), (3), (5), (7), (9) of Fig. 8.

A systematic study of the dependence of the energy barriers on the stripe geometry and material wettability is out of the scope of the present work; however it is worth mentioning that the presented methodology can be straightforwardly paired with optimization algorithms in order to design surface structures that resist impalement transitions or facilitate switching between certain wetting states.

4. Summary and Conclusions

The aim of this work is to develop a computationally efficient methodology, which simulates wetting of rough solids by droplets featuring multiple three phase contact lines. In particular, we solve the Young Laplace equation augmented with a disjoining pressure term, in order to incorporate the liquid/solid interactions in the vicinity of the solid surface. This methodology enabled us to bypass the implementation of the Young's contact angle, $\theta_Y$, boundary condition, since the liquid/vapor and the liquid/solid interfaces are treated in a unified framework. The Young's contact angle emerges implicitly from the combined action of liquid/vapor (surface tension) and liquid/solid interactions, which are parameterized through a wetting parameter, $w^{LS}$.

To our knowledge, this is the first time that a one-dimensional continuum level model is able to predict Cassie-Baxter wetting states and the accompanying transitions to Wenzel ones by simulating the entire droplet profile and not a unit cell. The main advantage of the proposed methodology is the ability to compute multiple and reconfigurable three phase contact lines by accounting for microscale interactions in the Young Laplace equation and treat the liquid/vapor and the liquid/solid interface by a unified way. Alternative continuum-level methodologies include tracking interface techniques (e.g. level-set [42, 43]) and surface evolving methods for energy minimization ("Surface evolver" software [44-46]). The proposed augmented YL equation is advantageous over these methodologies: e.g. level-set requires an additional space dimension to track the droplet interface as well energy minimization using the "Surface evolver" needs to pre-define the number of TPLs limiting the attainable equilibrium droplet solutions. Molecular Dynamics [19, 20] or Lattice-Boltzmann [21, 22] simulations implemented for the same task demand prohibitively higher computational power especially when real millimeter size droplets are studied.

The augmented YL equation is validated against the conventional YL equation for droplets wetting simple structured solid surfaces. Moreover, a comparison with a conceptually different approach, and in particular with LB simulations, showed satisfactory agreement between the two different methodologies. We also demonstrated the efficiency of the proposed methodology to compute all admissible wetting states on a stripe-patterned solid surface. By performing parametric continuation, we compute multiple co-existing states, and study their relative stability. By computing the free energy difference between stable and unstable wetting states we calculate the minimum energy required to induce certain wetting transitions.

The study of the dependence of the computed energy barriers on various geometric features of a rough surface would strongly impact the design of surfaces with addressable wetting properties in terms of high resistance in impalement transitions or efficient switching between wetting states.

## 5. Acknowledgements

The authors kindly acknowledge funding from the European Research Council under the Europeans Community's Seventh Framework Programme (FP7/2007–2013)/ERC grant agreement no. [240710]. We would also like to greatly acknowledge Prof. I. Kevrekidis and Dr. G. Kokkoris for their valuable and inspiring suggestions.

## Appendix

A1. Eikonal equation

The disjoining pressure, $p^{LS}$, is a function of Euclidean distance from the solid boundary. For structured surfaces, this distance is computed solving the Eikonal equation, which reads:

$$|\nabla \delta(x,y)| = 1, \ x,y \in \Phi, \tag{A1-1}$$

$$\delta(x,y) = 0, \ x,y \in \partial S, \tag{A1-2}$$

where $\Phi$ corresponds to the two-dimensional computational domain and $\partial S$ to the solid surface boundary. The equation is solved with the vanishing viscosity method [47]. The distance, $\delta$, of a point located at the droplet surface, from the solid boundary, is interpolated from the solution of the Eikonal equation. Thus, the Eikonal equation is solved only once for a particular solid surface geometry.

A2. Circular fitting

Circular fitting of the droplet surface is performed using the least square approximation (see embedded figure of Fig. 2) at a level, beyond which the liquid/solid interactions are of negligible magnitude, and a constant curvature of the droplet is to be expected. The Young's contact angle is evaluated from:

$$\theta_Y = acos\left(1 - \frac{h}{R}\right), \tag{A2-1}$$

where $R$ is the radius of fitting circle and $h$ the droplet height.